\documentclass{aastex631}
\pdfoutput=1

\usepackage{amssymb,amsmath}
\usepackage{soul}

\usepackage{tikz}
\usetikzlibrary{shapes.geometric, arrows}

\tikzstyle{roundrect} = [rectangle, rounded corners, 
minimum width=3cm, 
minimum height=1cm,
text width=3.8cm,
text centered, 
draw=black, 
fill=orange!15]

\tikzstyle{arrow} = [thick,->,>=stealth]

\defcitealias{Brandt_2023}{Paper I}

\begin{document}

\title{Likelihood-Based Jump Detection and Cosmic Ray Rejection for Detectors Read Out Up-the-Ramp}

\author[0000-0003-2630-8073]{Timothy D.~Brandt}
\affiliation{Space Telescope Science Institute \\ 3700 San Martin Drive \\ Baltimore, MD 21218, USA}
\affiliation{Department of Physics, University of California, Santa Barbara \\ Broida Hall \\ Santa Barbara, CA, 93106, USA}

\begin{abstract}

This paper implements likelihood-based jump detection for detectors read out up-the-ramp, using the entire set of reads to compute likelihoods.  The approach compares the $\chi^2$ value of a fit with and without a jump for every possible jump location.  I show that this approach can be substantially more sensitive than one that only uses the difference between sequential groups of reads, especially for long ramps and for jumps that occur in the middle of a group of reads.  It can also be implemented for a computational cost that is linear in the number of resultants.  I provide and describe a pure Python implementation that can process a 10-resultant ramp on a $4096 \times 4096$ detector in $\approx$20 seconds, including iterative cosmic ray detection and removal, on a single core of a 2020 Macbook Air.  This Python implementation, together with tests and a tutorial notebook, are available at \url{https://github.com/t-brandt/fitramp}.  I also provide tests and demonstrations of the full ramp fitting and cosmic ray rejection approach on data from JWST.

\end{abstract}


\section{Introduction} \label{sec:intro}

Cosmic ray hits are a common problem in astronomical images. 
They inevitably cause data loss if identified, and can corrupt data if not properly identified.  In a charge coupled device (CCD), each pixel accumulates electrons that are read out once.  If an energetic cosmic ray hits a pixel during an exposure, that exposure for that pixel is contaminated beyond recovery. 
All that can be done is to identify the cosmic ray by the spatial structure of the counts on the array \citep{vanDokkum_2001,Pych_2004}.  The data must then be discarded or, if necessary, interpolated over \citep{Zhang+Bloom_2020,Zhang+Brandt_2021}.  

For an infrared detector that can be read out nondestructively, the story is potentially much more promising than for CCDs.  Cosmic rays might affect only a small part of a ramp.  In that case, the reads before and the reads after the cosmic ray might both be able to constrain that pixel's photon rate.  Cosmic ray identification and rejection has been a part of nondestructive ramp fitting approaches for decades \citep[e.g.][]{Fixsen+Offenberg+Hanisch+etal_2000,Offenberg+Fixsen+Rauscher+etal_2001}.  \cite{Anderson+Gordon_2011} compared three different statistical tests for a jump in a pixel's counts.  Once a cosmic ray is identified, the contaminated portion of the ramp can be discarded, with the remainder being kept.

This paper is the second in a two-paper sequence on ramp fitting and jump detection.  \cite{Brandt_2023}, hereafter \citetalias{Brandt_2023}, presented a likelihood-based fit to a nondestructive ramp.  The present paper applies this approach to search for differences in groups of reads that indicate a jump, e.g., from a cosmic ray.  By using the entire ramp in a likelihood-based formalism, this new approach can make more efficient use of the data at hand.  

I organize the paper as follows.  In Section \ref{sec:ramp_overview}, I summarize the results of \citetalias{Brandt_2023} that are needed here.  Section \ref{sec:jumpdetect} presents the mathematical approach to searching for jumps and proposes a practical algorithm.  Section \ref{sec:jumpdetect_demo} includes a series of sensitivity tests on synthetic data, comparing the results of the new algorithm to existing approaches.  Section \ref{sec:implementation} briefly summarizes the implementation; the algorithms are included with the same package as those presented in \citetalias{Brandt_2023}.  Section \ref{sec:examples} applies the full ramp fit and jump detection approach to two JWST data sets, comparing the results to the processed data available on the MAST archive as of this writing.  I conclude with Section \ref{sec:conclusions}.

\section{Overview of $\chi^2$ Ramp Fitting}
\label{sec:ramp_overview}

\begin{figure}
    \centering\includegraphics[width=0.5\textwidth]{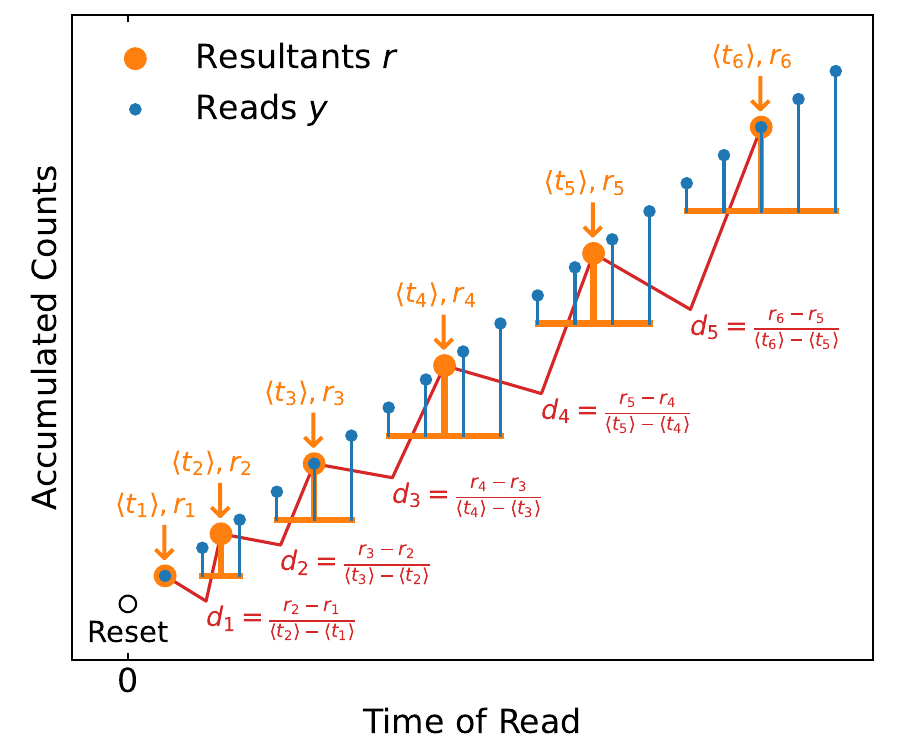}
    \caption{{Illustration of the process of averaging nondestructive reads (blue points) into ``groups'' (JWST terminology) or ``resultants'' (Roman terminology, used in this paper).  Resultants are labeled as $r_1, r_2, \ldots$ and denoted by orange points; the resultants comprise the ramp.  The group fitting operates in the space of scaled resultant differences, denoted by $d_1, d_2, \ldots$, scaled by the difference in mean time between resultants.  Each $d_i$ is an unbiased estimator of the count rate.  Figure reproduced from \citetalias{Brandt_2023}. }\label{fig:ramp_cartoon}}
\end{figure}

\citetalias{Brandt_2023} derived the maximum likelihood fit to a nondestructive ramp in the presence of read and photon noise.  Reads can be averaged together before they are made available for further processing; these are known as either ``groups'' (e.g.~by JWST) or as ``resultants'' (e.g.~by Roman).  In the remainder of this paper I will refer to them as resultants.  {Figure \ref{fig:ramp_cartoon}, reproduced from \citetalias{Brandt_2023}, illustrates the averaging of reads into resultants that ultimately produces the ramp to be fit.  The ramp fitting of \citetalias{Brandt_2023}, and the jump detection of this paper, operate on the space of resultant differences scaled by the mean time between resultants; these are denoted by $d_1, d_2, \ldots$ in Figure \ref{fig:ramp_cartoon}.}  The formalism of \citetalias{Brandt_2023} applies to any readout scheme and, crucially, can accommodate any number of unused resultant differences.  A resultant difference could be unused because of saturation or because of a jump between resultants or within a resultant--due to a reset, a cosmic ray, or some other fluctuation.  

\citetalias{Brandt_2023} recasts the problem of fitting a ramp from dealing with accumulated counts to dealing with count differences.  This approach makes the covariance matrix of the resultant differences tridiagonal: the difference between two consecutive resultants shares reads and photons only with the difference between the next two and the previous two consecutive resultants.  A tridiagonal covariance matrix admits an analytic inverse, and the equations for $\chi^2$ can be reorganized and simplified. 

The equations of \citetalias{Brandt_2023} are in closed form and have a computational cost that is linear in the number of groups of reads.  They yield the maximum likelihood count rate together with the $\chi^2$ value of a fit to the nondestructive ramp. This $\chi^2$ value, interpreted as a log likelihood, forms the basis of the jump detection algorithm described in the remainder of this paper.

\section{Cosmic Ray Detection via Hypothesis Testing} \label{sec:jumpdetect}

The key result of \citetalias{Brandt_2023} for the present discussion is a closed form for the standard $\chi^2$ statistic for a ramp with the omission of an arbitrary number of resultant differences.  
The availability of $\chi^2$ enables hypothesis testing and provides a straightforward goodness of fit measurement.  In this section, I will restrict the discussion to the case where the reset value is not being fit, i.e., where $d_1$ in Sections 3 and 4 of \citetalias{Brandt_2023} refers to the scaled difference between the first and second resultants.  In the case of an informative prior on the reset value the algebra would be more complicated; I do not currently treat a fit with a pedestal together with jump detection.  The fit to a ramp has one free parameter (the count rate), so the number of degrees of freedom is the number of resultant differences minus one.  The survival function of this $\chi^2$ distribution can provide a natural flag for bad fits.  

The availability of $\chi^2$ also enables hypothesis testing through likelihood ratios.  In this section I develop one application of this hypothesis testing: searching for cosmic rays that manifest as instantaneous jumps in the counts at a given pixel.  The approach is equivalent to the $y$-intercept approach explored by \cite{Anderson+Gordon_2011} but the algorithms provided here are more efficient.

\subsection{Cosmic Ray Detection Between Resultants} 
\label{subsec:omitone}

A jump in a ramp is a sudden increase in the number of counts.  When fitting a line to the accumulated counts, accounting for the jump requires fitting an offset between two ramps of identical slope.  In the framework outlined here, it means omitting the resultant difference that is corrupted by the jump.  If a jump happens between two reads of the same resultant, two differences---both of the ones that use the contaminated group---will need to be discarded.  This can either be done by fitting one additional free parameter per resultant difference to be discarded, or by omitting these resultant differences (which alters the covariance matrix).  I will take the former approach here.  This subsection will treat the case where a jump occurs between resultants; the following subsection will treat the case where a jump occurs within a resultant.

In this section I work out the best-fit slope, its uncertainty, and the best-fit $\chi^2$ when omitting one resultant difference by giving it an additional free parameter.  I will assume that the resultant difference to be omitted is $j$ and use ${\bf j}$ to denote a vector that is zero except for element $j$, where it is one.  The new expression for $\chi^2$ is
\begin{align}
    \chi^2 &= \left({\bf d} - a_j {\bf 1} - b_j {\bf j}\right)^T {\bf C}^{-1} \left({\bf d} - a_j {\bf 1} - b_j {\bf j}\right) \nonumber \\
    &= {\bf d}^T {\bf C}^{-1} {\bf d}
    + a_j^2 \cdot {\bf 1}^T {\bf C}^{-1} {\bf 1}
    + b_j^2 \cdot {\bf j}^T {\bf C}^{-1} {\bf j}
    - 2b_j \cdot {\bf d}^T {\bf C}^{-1} {\bf j}
    - 2a_j \cdot {\bf d}^T {\bf C}^{-1} {\bf 1}
    + 2a_j b_j \cdot {\bf j}^T {\bf C}^{-1} {\bf 1} .
\end{align}
This closely matches Equation (68) from \citetalias{Brandt_2023}, and I take the same approach to minimize $\chi^2$ and compute the covariance matrix.  I generalize three terms defined in Section 4.1 of \citetalias{Brandt_2023} to any read $j$:
\begin{equation}
    {\bf j}^T {\bf C}^{-1} {\bf j} = {C}^{-1}_{jj} = \frac{\theta_{j-1} \phi_{j+1}}{\theta_n},
\end{equation}
\begin{equation}
    {\bf j}^T {\bf C}^{-1} {\bf 1} \equiv {\cal C}'_j = \frac{\left(-1\right)^j}{\theta_n} \left( \phi_{j+1} \Theta_j +  \theta_{j-1} \Phi_j \right) ,
\end{equation}
and
\begin{equation}
    {\bf d}^T {\bf C}^{-1} {\bf j} \equiv {\cal B}'_j = \frac{\left(-1\right)^j}{\theta_n} \left( \phi_{j+1} (\Theta D)_j +  \theta_{j-1} (\Phi D)_j \right) .
\end{equation}
This allows me to write the expression for $\chi^2$ as 
\begin{equation}
    \chi^2 = {\cal A} + a_j^2 {\cal C} + b_j^2 C^{-1}_{jj} - 2b_j {\cal B}'_j - 2a_j {\cal B} + 2a_j b_j {\cal C}'_j.
    \label{eq:chi2_omitone}
\end{equation}
All sums only need to be computed once for the ramp, not once per $j$.  As a result, computing $\chi^2$ at all $j$, i.e., at all possible locations of a jump in the counts, has a cost that is linear in the number of resultants.  

I will now derive the best-fit $\chi^2$, the corresponding slope, and its uncertainty.  By computing the best-fit $\chi^2$ at all possible jump locations we can perform rigorous hypothesis testing using all available information.  The first step is to differentiate $\chi^2$ and set the result equal to zero:
\begin{align}
    \frac{\partial \chi^2}{\partial a_j} &= 0 = 2a_j {\cal C} - 2 {\cal B} + 2b_j {\cal C}'_j \\
    \frac{\partial \chi^2}{\partial b_j} &= 0 = 2b_j C^{-1}_{jj} - 2 {\cal B}'_j + 2a_j {\cal C}'_j .
\end{align}
This yields
\begin{align}
    a_j &= \frac{C^{-1}_{jj} {\cal B} - {\cal B}'_j {\cal C}'_j}{{\cal C} C^{-1}_{jj} - \left({\cal C}'_j\right)^2} \\
    b_j &= \frac{{\cal B}'_j - a_j {\cal C}'_j}{C^{-1}_{jj}} .
\end{align}
These expressions can be efficiently and simultaneously computed at all possible $j$ values.  They may then be substituted into Equation \eqref{eq:chi2_omitone} to derive the best $\chi^2$ possible by omitting each possible resultant difference.

The final step is to derive the standard uncertainty on the best-fit count rate.  To do this I will first derive the {inverse} covariance matrix of $a_j$ and $b_j$ as
\begin{align}
    {\bf C}^{-1}(a_j, b_j) &= 
    \begin{bmatrix}
        \frac{1}{2}\frac{\partial^2 \chi^2}{\partial a_j^2} & \frac{1}{2}\frac{\partial^2 \chi^2}{\partial a_j \partial b_j} \\
        \frac{1}{2}\frac{\partial^2 \chi^2}{\partial a_j \partial b_j} & \frac{1}{2}\frac{\partial^2 \chi^2}{\partial b_j^2}
    \end{bmatrix} \nonumber \\
    &= \begin{bmatrix}
        {\cal C} & {\cal C}'_j \\
        {\cal C}'_j & C^{-1}_{jj}
    \end{bmatrix} .
\end{align}
This finally gives the standard uncertainty on $a_j$ as
\begin{equation}
\sigma^2(a_j) = \frac{C^{-1}_{jj}}{{\cal C}C^{-1}_{jj} - ({\cal C}'_j)^2} .
\end{equation}

The value of $b_j$ may also be useful, e.g., if fitting for the morphology of a cosmic ray hit to correct pixels where the hit induced only a small perturbation to the ramp.  In this case, the uncertainty in $b_j$ may also be of value; it is given by
\begin{equation}
    \sigma^2(b_j) = \frac{\cal C}{{\cal C}C^{-1}_{jj} - ({\cal C}'_j)^2} .
\end{equation}
If there is a large jump at resultant difference $j$, then $\chi^2_j$ will be much less than the $\chi^2$ value including all resultants, and $a_j$ will be the maximum likelihood estimate of the count rate accounting for the jump.

\subsection{Cosmic Ray Detection Within a Resultant} \label{subsec:omittwo}

The previous subsection treated jumps that occur between resultants so that only one difference must be omitted.  In this subsection I derive the formulas for the best-fit count rate, its uncertainty, and the $\chi^2$ of the fit when omitting two adjacent resultant differences.  This is necessary when a jump occurs between reads of the same group or resultant. 
 The expression for $\chi^2$ becomes
\begin{align}
    \chi^2 = \left({\bf d} - a {\bf 1} - b {\bf j} - c {\bf k} \right)^T {\bf C}^{-1} \left({\bf d} - a {\bf 1} - b {\bf j} - c {\bf k} \right) 
\end{align}
where ${\bf j}$ is a vector that is zero except for the $j$ element which is one, and ${\bf k}$ is a vector that is zero except for the $k$ element which is one, with $k=j+1$.  This corresponds to freely fitting for the actual resultant differences $j$ and $j+1$ so that they have no influence on the count rate.  I can multiply this out to obtain
\begin{align}
    \chi^2 &= {\bf d}^T {\bf C}^{-1} {\bf d} + a^2 \cdot {\bf 1}^T {\bf C}^{-1} {\bf 1}
    + b^2 \cdot {\bf j}^T {\bf C}^{-1} {\bf j}
    + c^2 \cdot {\bf k}^T {\bf C}^{-1} {\bf k}
    - 2a \cdot {\bf d}^T {\bf C}^{-1} {\bf 1}
    - 2b \cdot {\bf d}^T {\bf C}^{-1} {\bf j}
    \nonumber \\
    &\qquad\qquad - 2c \cdot {\bf d}^T {\bf C}^{-1} {\bf k}
    + 2ab \cdot {\bf 1}^T {\bf C}^{-1} {\bf j}
    + 2ac \cdot {\bf 1}^T {\bf C}^{-1} {\bf k}
    + 2bc \cdot {\bf j}^T {\bf C}^{-1} {\bf k} \\
    &= {\cal A} + a^2 {\cal C} + b^2 C^{-1}_{jj} + c^2 C^{-1}_{kk} -2a {\cal B} - 2b {\cal B}'_j - 2c {\cal B}'_k + 2ab {\cal C}'_j + 2ac {\cal C}'_k + 2bc C^{-1}_{jk} .
\end{align}
All of these have already been computed apart from
\begin{align}
    C^{-1}_{jk} = C^{-1}_{j,j+1} = -\frac{\beta_j \theta_{j-1}\phi_{j+2}}{\theta_n} .
\end{align}
Differentiating $\chi^2$ to find the best-fit $a$, $b$, and $c$, I have
\begin{align}
    \frac{1}{2} \frac{\partial \chi^2}{\partial a} &= 0 = a {\cal C} + b {\cal C}'_j + c {\cal C}'_k - {\cal B} \\
    \frac{1}{2} \frac{\partial \chi^2}{\partial b} &= 0 = a {\cal C}'_j + b C^{-1}_{jj} + c C^{-1}_{jk} - {\cal B}'_j \\
    \frac{1}{2} \frac{\partial \chi^2}{\partial c} &= 0 = a {\cal C}'_k + b C^{-1}_{jk} + c C^{-1}_{kk} - {\cal B}'_k
\end{align}
which can be solved by hand to yield
\begin{align}
    a &= \frac{\cal B}{\cal C} - b \frac{{\cal C}'_j}{\cal C} - c \frac{{\cal C}'_k}{\cal C} \\
    b &= \frac{{\cal B}{\cal C}'_j - {\cal B}'_j {\cal C}}{\left({\cal C}'_j\right)^2 - {\cal C}C^{-1}_{jj}} - c \left( \frac{{\cal C}'_j {\cal C}'_k - C^{-1}_{jk} {\cal C}}{\left({\cal C}'_j\right)^2 - {\cal C}C^{-1}_{jj}}\right) \\
    c &= \left( \frac{{\cal B}{\cal C}'_j - {\cal B}'_j {\cal C}}{\left({\cal C}'_j\right)^2 - {\cal C}C^{-1}_{jj}} - \frac{{\cal B}{\cal C}'_k - {\cal B}'_k {\cal C}}{{\cal C}'_j {\cal C}'_k - {\cal C} C^{-1}_{jk}} \right) 
    \left( \frac{{\cal C}'_j {\cal C}'_k - {\cal C} C^{-1}_{jk}}{\left({\cal C}'_j\right)^2 - {\cal C}C^{-1}_{jj}} - \frac{\left({\cal C}'_k\right)^2 - {\cal C} C^{-1}_{kk}}{{\cal C}'_j {\cal C}'_k - {\cal C} C^{-1}_{jk}} \right)^{-1} .
\end{align}

The inverse of the covariance matrix for $a$, $b$, and $c$ is
\begin{equation}
    {\bf C}^{-1}(a, b, c) = 
    \begin{bmatrix}
    {\cal C} & {\cal C}'_j & {\cal C}'_k \\
    {\cal C}'_j & C^{-1}_{jj} & C^{-1}_{jk} \\
    {\cal C}'_k & C^{-1}_{jk} & C^{-1}_{kk}
    \end{bmatrix}
\end{equation}
so the variance on $a$ is
\begin{equation}
    \sigma^2(a) = \frac{C^{-1}_{jj} C^{-1}_{kk} - \left(C^{-1}_{jk}\right)^2}{{\cal C}\left(C^{-1}_{jj} C^{-1}_{kk} - \left(C^{-1}_{jk}\right)^2 \right) - {\cal C}'_j \left({\cal C}'_j C^{-1}_{kk} - C^{-1}_{jk} {\cal C}'_k \right) + {\cal C}'_k \left({\cal C}'_j C^{-1}_{jk} - C^{-1}_{jj} {\cal C}'_k\right)} .
\end{equation}

These expressions are more unwieldy than in the case of a jump between resultants, but their computational complexity remains linear in the number of resultants.  A large jump within a resultant would see the $\chi^2$ value much reduced when omitting both differences including that resultant; the corresponding value of $a$ would again be the maximum likelihood count rate.

\subsection{A Practical Approach for Multiple Jumps}
\label{sec:jumpdetect_algorithm}

The preceding subsections showed how to compute $\chi^2$ when leaving out a single resultant difference or a pair of differences.  The approach in Section \ref{subsec:omitone} is suitable for a jump that occurred between resultants, which is always the case when each resultant is a single read.  If a resultant consists of more than one read, a cosmic ray could arrive in the middle of the resultant.  In this case two resultant differences---both of the ones that contain the contaminated resultant---should be discarded using the formulae in Section \ref{subsec:omittwo}.

A real ramp could have more than one jump, and its resultants could contain a mixture of single reads and multiple reads.  In this case we need an iterative approach, and we need to treat resultants differently.  Here I propose a cosmic ray flagging and removal algorithm that satisfies both criteria.

The first step in my proposed approach is to treat single read resultants and multiple read resultants differently.  It will search for cosmic ray hits between single read resultants (potentially discarding only one difference), but will search both within and between resultants when they contain multiple reads. 
In all cases, a fit excluding single resultant differences or pairs of resultant differences will improve the $\chi^2$ value of the fit because it removes one or two constraints.  If this $\chi^2$ improvement exceeds a user-defined threshold for any resultant in a ramp, then a single difference {or pair of differences} will be discarded.  The discarded measurement will be the resultant difference or pair of differences whose $\chi^2$ improvement most exceeds the user-specified threshold.  These thresholds should differ when discarding one vs.~two resultant differences.  The lower threshold when discarding one resultant difference allows the approach to identify jumps between resultants even if those resultants consist of multiple {reads}.  If a jump occurs between resultants, every resultant should be usable and fit well within a partial ramp.  The best $\chi^2$ value discarding the difference with the jump should be {almost as good as} the best $\chi^2$ value when also discarding the subsequent (uncontaminated) difference. 

Once a difference or pair of differences has been identified as hosting a jump, it can be masked using the approach described in Section 4.2 of \citetalias{Brandt_2023}.  The algorithm outlined below can then proceed iteratively, with the process continuing until no additional resultant differences are flagged.

\begin{figure}
\begin{tikzpicture}[node distance=2cm]

\node (start) [roundrect] {Fit the ramp using all valid, unsaturated resultant differences};
\node (refit) [roundrect, below of=start] {Re-fit the ramp omitting each resultant difference and/or each pair of differences in turn};
\node (check_chisq) [roundrect, below of=refit, yshift=-0.2cm] {Does any new omission improve $\chi^2$ by more than the adopted threshold?};

\node (mask) [roundrect, right of=check_chisq, xshift=2.8cm] {Mask the resultant difference or pair of differences whose omission most improves $\chi^2$};

\node (adopt) [roundrect, below of=check_chisq, yshift=-0.1cm] {Adopt the mask from all previous\\ iterations combined};
\node (debias) [roundrect, below of=adopt, yshift=0.2cm] {Re-fit the ramp one more time with this new mask to remove bias};
\draw [arrow] (start) -- (refit);
\draw [arrow] (refit) -- (check_chisq);
\draw [arrow] (check_chisq) -- node[anchor=south] {yes} (mask);
\draw [arrow] (check_chisq) -- node[anchor=west] {no} (adopt);
\draw [arrow] (mask) |- (refit);
\draw [arrow] (adopt) -- (debias);
\end{tikzpicture}
\caption{Flow chart of the proposed jump detection and masking algorithm described in Section \ref{sec:jumpdetect_algorithm}.  \label{fig:flowchart}}
\end{figure}
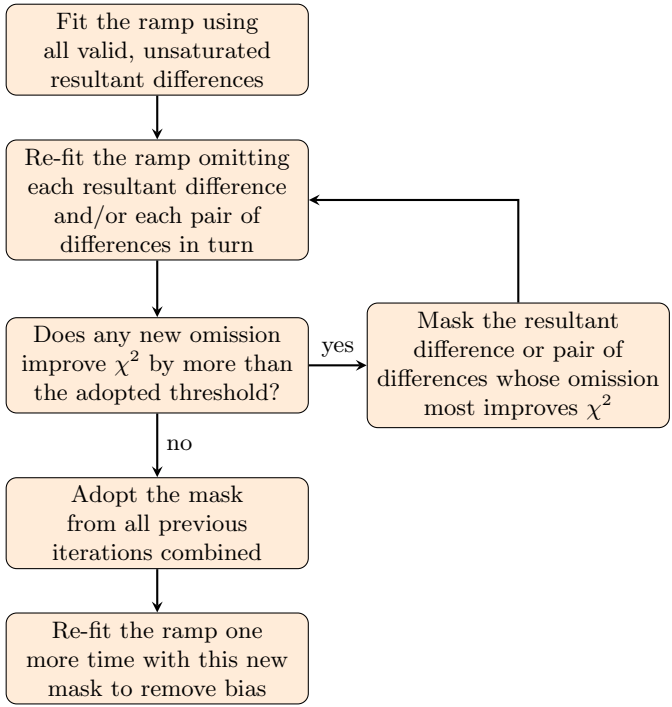

A step-by-step approach of the algorithm above follows.  Figure \ref{fig:flowchart} presents a flowchart summarizing the algorithm.
\begin{enumerate}
    \item Using the method in Sections \ref{subsec:omitone} and \ref{subsec:omittwo}, identify the resultant difference or pair of differences most likely to hold a jump, i.e., the one whose omission most improves $\chi^2$ over the jump detection threshold.  For a difference of single read resultants only one resultant difference will be omitted, while for a resultant containing two or more reads either one or two adjacent differences can be omitted.  For this step, I first estimate the covariance matrix using the median of the resultant differences for the count rate.  If the estimated photon count rate is negative {I} set it to zero.
    \item Determine whether the improvement in $\chi^2$ from omitting this resultant difference, indexed by $j$, exceeds a user-specified significance threshold.  When comparing the possibility of discarding a single resultant difference vs.~two differences, the one that offers the largest $\chi^2$ improvement over the relevant user-specified threshold will be chosen.
    \item If the resultant difference $j$ or pair of differences $j,j+1$ that offers the greatest $\chi^2$ improvement exceeds the user-specified threshold, discard it from the analysis using the approach of Section 4.2 in \citetalias{Brandt_2023}. 
    \item Return to Step 1 and iterate until no resultant difference or differences exceeds the adopted significance threshold, or until the ramp contains two or fewer resultant differences.  If there are only two resultant differences remaining there is no way to tell which one is correct and which contains a jump.  In this case, the entire ramp is corrupted.
    \item After discarding all flagged resultant differences, adopt the resulting count rate, its uncertainty, and $\chi^2$ value for that pixel.  
\end{enumerate}
For Step 3, we must also explicitly set $C^{-1}_{jj}=0$ in addition to the methodology of Section 4.2 of \citetalias{Brandt_2023}.  This will result in expressions containing $0/0$ for $a$, $b$, and/or $c$ if any of the relevant resultant differences are ignored in a later iteration.  {This situation arises because the quantities $a$, $b$, and/or $c$ are most naturally computed simultaneously for all resultant differences; the cost is small compared with the cost of fitting the ramp in the first place.  It is a matter taste where and how to account for the possibility of checking a resultant difference twice.  Instead of using logical operations at the beginning to test each difference or pair of differences to see if it should be checked, I implement an equivalent check later by replacing data whose calculation involved 0/0.}  In the case that a calculation does return 0/0, a given resultant difference was already excluded in a calculation with one fewer resultant difference omitted.  We can therefore fill in these values with the corresponding count rates, $\chi^2$ values, and uncertainties from our prior calculation that did not doubly omit a resultant difference.

After applying this sequence of steps, the user can re-estimate the covariance matrix one final time with the count rates from Step (5) to obtain the final count rates and uncertainties.  This removes biases in the count rates to first order as discussed in Section 5 of \citetalias{Brandt_2023}.  

The approach outlined above requires no modification of the core equations of \citetalias{Brandt_2023}.  It does require running the algorithm several times for those pixels with more than one jump, with a total computational cost a factor of a few times the cost of fitting a single ramp once.  It is also well-suited to taking ramps that have already had most jumps identified and checking for any additional, lower signal-to-noise ratio jumps.

The equations in this section give the best-fit $\chi^2$, the best-fit count rate and its standard uncertainty leaving out any individual resultant difference or pair of adjacent differences.   The results for all possible omitted resultants may be computed for a computational cost linear in the number of resultants, i.e., for a similar cost to computing the best-fit count rate in the first place. This renders a full, likelihood-based cosmic ray detection algorithm computationally straightforward.  Section \ref{sec:implementation} discusses the computational cost in more detail.

\section{Demonstration of Jump Fitting With Synthetic Data} \label{sec:jumpdetect_demo}

The algorithm outlined in the preceding section can identify jumps by the $\chi^2$ improvement yielded by excluding one or two resultant differences at a time.  Whether one or two differences are excluded depends on whether either resultant consisted of two or more reads.  In this section I show examples of the performance this technique enables.  I will compare to an algorithm restricted to looking at the differences between adjacent resultants and flagging them if they exceed a threshold.  For my example here I use a threshold of 4.5 times the root variance of a resultant difference. 
If a resultant difference exceeds the median resultant difference over the ramp by at least 4.5 times the root variance of a single difference, then we discard that difference.  My choice of the median rather than the mean for the baseline resultant difference is to limit the bias in the reference count rate from the presence of a jump.  The mean is more sensitive to outliers introduced by the jumps themselves; this comes with a corresponding loss of sensitivity.  Repeating the analysis of the following section with means rather than medians degrades sensitivity to jumps, typically by a few percent.

I will show two cases separately.  First, I will consider a long ramp composed of individual reads.  In this case one resultant difference may be left out at a time.  Second, I will consider a shorter ramp in which each resultant consists of several reads.  

\subsection{Single-Read Resultants}

\begin{figure}
    \includegraphics[width=0.5\textwidth]{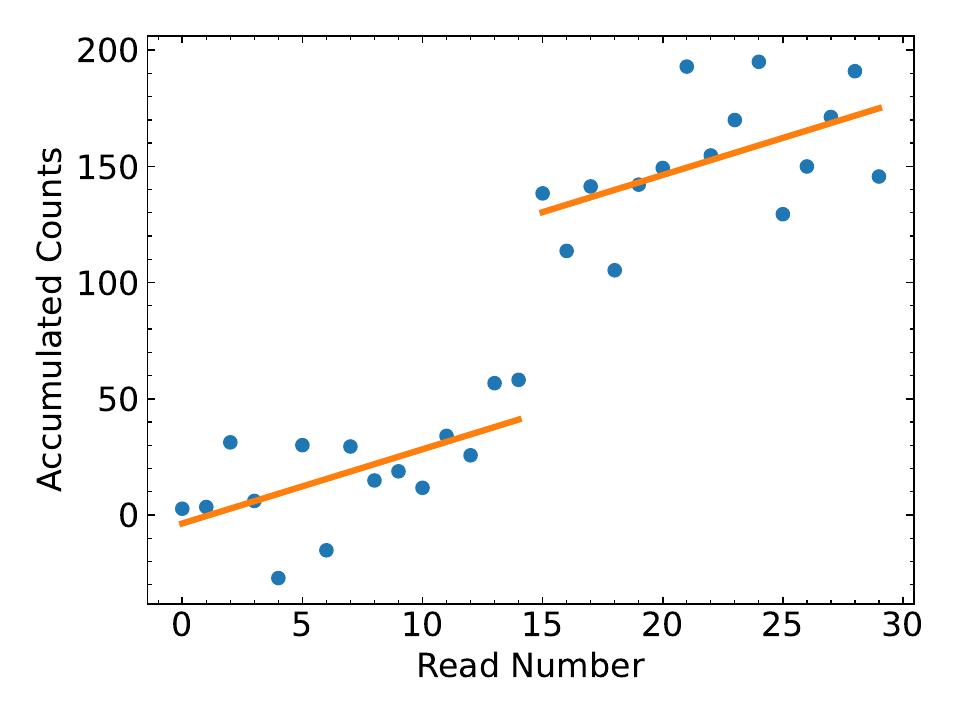}
    \includegraphics[width=0.5\textwidth]{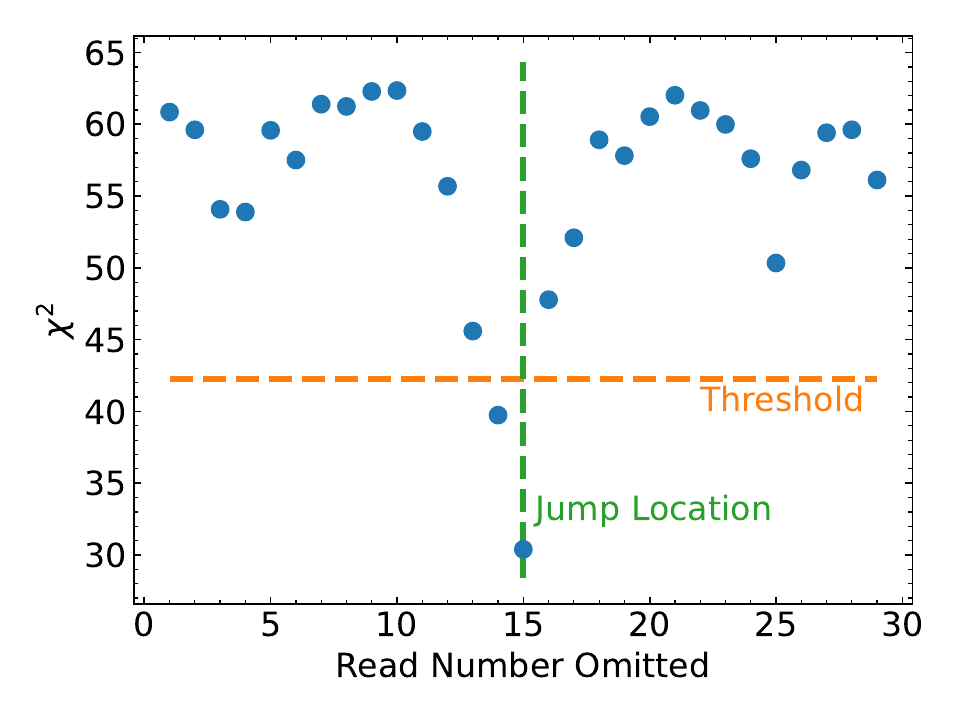}
    \caption{Demonstration of the likelihood-based jump detection algorithm.  Left panel: a ramp showing the accumulated counts with the best-fit slope also shown.  The ramp has a single-read read noise of 20\,e$^-$ and a true count rate of 20\,e$^-$/read.  The slope was fitted to the read differences using the proper covariance matrix; the offsets are approximate.  Right panel: $\chi^2$ of the fit to the read differences as a function of the read pair omitted.  The orange line marked ``Threshold'' denotes an improvement to $\chi^2$ corresponding to 4.5$\sigma$ significance.  The green dashed line shows where I added a jump; the jump is 3 times the uncertainty in the single read difference (CDS read noise plus photon noise added in quadrature).  Omitting the read with the jump offers the maximal improvement in $\chi^2$, and results in a formally good $\chi^2$ of 30.4 for 30 reads: 29 read differences with one not used and one fitted parameter, for 27 degrees of freedom.}
    \label{fig:jump_demo}
\end{figure}

I will begin by illustrating the likelihood-based approach on a 30-read ramp in which I have added a discrete jump at the 15th read (indexed from zero).  The ramp has with a true count rate of 20\,e$^-$/read and a single-read read noise of 20\,e$^-$.  Figure \ref{fig:jump_demo} shows the accumulated counts in the left panel, and the $\chi^2$ value of the best-fit line as a function of the read difference that is left out.  There is a large improvement to $\chi^2$ when leaving out the 15th read difference, exactly the one that in which I added a jump.  The cumulative distribution function of the $\chi^2$ distribution with one degree of freedom corresponds to the usual significance thresholds for a Gaussian distribution, e.g., 68.3\% of the probability is within $1\sigma$, or $\Delta \chi^2 = 1^2$.  In Figure \ref{fig:jump_demo} I have marked an improvement in $\chi^2$ of 20.25, which corresponds to 4.5$\sigma$.  If there is no jump and the statistical description of the errors is correct, such an improvement should occur by chance with a probability of 
\begin{equation}
{\rm erfc} \left(\frac{4.5}{\sqrt{2}}\right) \approx 7\times10^{-6} .
\end{equation}
If any point exceeds this threshold then the read difference that offers the maximal improvement to $\chi^2$ when left out of the fit is labeled as a jump.

To test the sensitivity of this approach to jumps we need to define a threshold in $\chi^2$ improvement over the single ramp at which we will label a jump significant.  If we want the equivalent of 4.5\,$\sigma$, we can use a $\Delta \chi^2$ threshold of 20.25; this enables a natural comparison with an algorithm that looks for individual read differences that are at least 4.5$\sigma$ discrepant with the average.  This significance threshold is shown in Figure \ref{fig:jump_demo}.  

\begin{figure}
\includegraphics[width=0.5\linewidth]{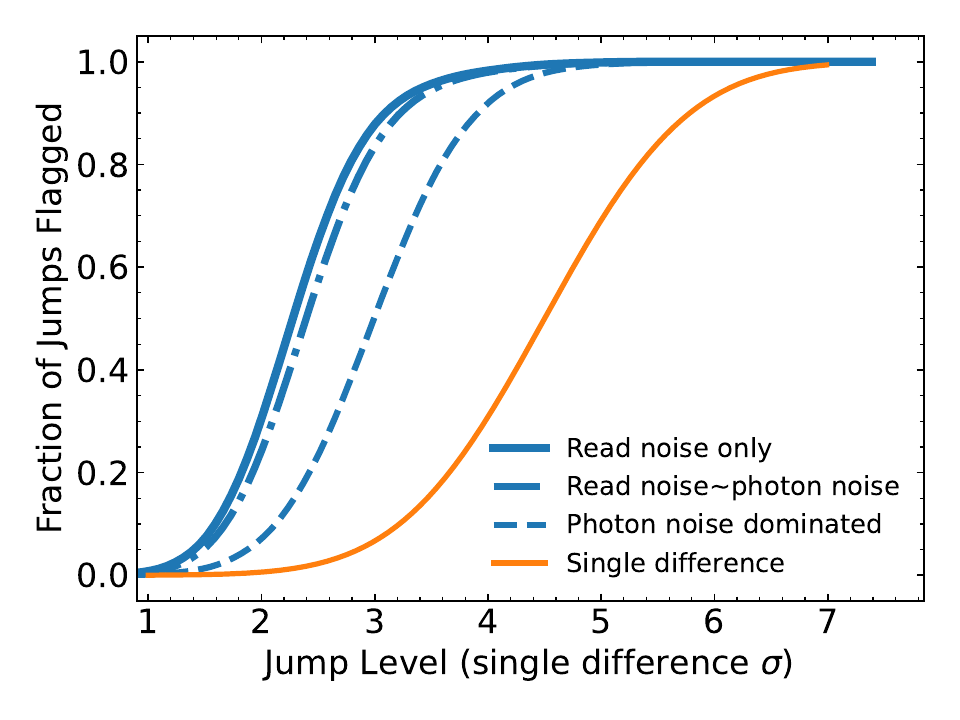}
\includegraphics[width=0.5\linewidth]{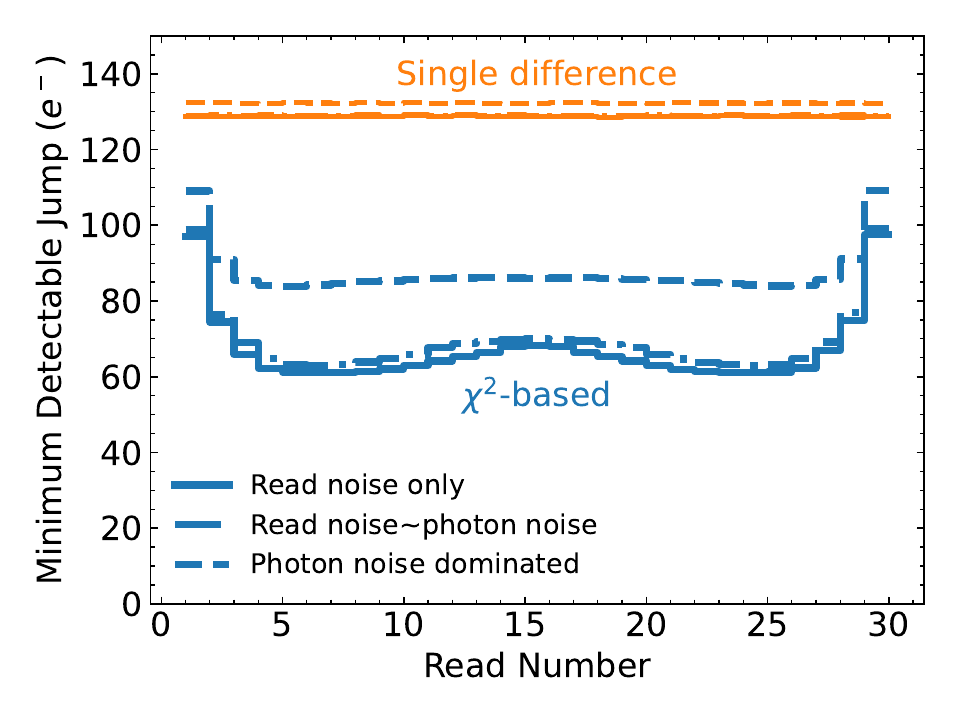}
    \caption{Left: comparison of jump detection efficiency on 30 read ramps with a read noise of 20 electrons/read.  The three sets of thicker lines use the likelihood approach with a threshold of a $\chi^2$ improvement of 20.25 when omitting a read; this is equivalent to 4.5$\sigma$.  Equal read and photon noise refers to the contributions over the ramp rather than in an individual difference (which is then dominated by read noise); in this case it is 4.{6}\,e$^-$/read.  The ``mostly photon noise'' case has a photon rate of 4{6}\,e$^-$/read; read noise is still important in individual differences between reads.  The thinner {orange} line shows the performance of a single difference between adjacent reads with a 4.5$\sigma$ threshold; {it is a Gaussian cumulative distribution function translated by $4.5\sigma$}. Right: the minimum jump for a 50\% detection probability as a function of its location within a 30 read ramp with the same read noise and count rates as the left panel.  Jumps near the middle of a ramp are easier to detect using a $\chi^2$-based approach because both sides of the jump may be used effectively to fit a slope.  }
    \label{fig:jumpdetection_compare}
\end{figure}

Figure \ref{fig:jumpdetection_compare} shows the results of a full sensitivity analysis.  With 30 reads and in the photon noise limit, the likelihood-based jump detection is approximately twice as sensitive (on average) as the single difference approach.  This change in sensitivity depends on where in the ramp the jump occurs.  If the jump occurs in the middle of the reads the likelihood approach offers a larger sensitivity gain.  This is also the location at which the jump would induce the maximum increase in the best-fit count rate.  With more than 30 reads the advantage of the likelihood-based approach over the single difference approach grows: it becomes a factor of $\approx$2.4 at 50 reads, and $\approx$3.3 at 100 reads (assuming read noise is still significant).  The single difference approach only reaches the sensitivity of the likelihood-based approach when photon noise dominates the uncertainty in individual read differences rather than just in the entire ramp.  This only occurs for 
{count rates (in e$^-$/read) significantly higher than the square of the read noise (in the same units), or $\gg$400\,e$^-$/read for a read noise of 20\,e$^-$.}  
The right panel of Figure \ref{fig:jumpdetection_compare} shows the jump that is detected 50\% of the time as a function of its position within a 30-read ramp.  The single difference approach is equally sensitive everywhere, while the $\chi^2$ approach is most sensitive near the middle of the ramp where the jump would induce the maximal bias on the inferred count rate.  All of the synthetic ramps used for these tests consist of 30 reads each with a read noise of 20\,e$^-$ per read.  The case of equal read and photon noise is a count rate of 4.{6}\,e$^-$/read; this gives equal contributions of photon and read noise in the overall ramp fit (though not in individual reads).  {Photon and read noise, for this purpose, are each calculated using the product of the weight vector's transpose, the read noise or photon noise covariance matrix, and the weight vector, as described by Equation (110) of \citetalias{Brandt_2023}. The weight vector is the optimal set of weights for combining the resultants into a slope, Equations (99) and (100) of \citetalias{Brandt_2023}.}

\subsection{Multiple read resultants}

I now turn to the case of a shorter ramp in which each resultant is composed of multiple reads.  In this case the likelihood-based algorithm will omit {either one or} two resultant differences at a time. {If omitting two resultant differences, both omissions include the one} that is being investigated for a jump.  For this example I will consider a ramp of ten resultants with a jump in one of the resultants other than the first and last one.  I will compute the sensitivity of the algorithm to a jump as a function of the time of the cosmic ray hit, for resultants each composed of six reads, and for resultants each composed of many reads (60 in my synthetic ramps).  {The sensitivity behaves differently depending on whether the jump occurs between resultants, or between two reads that are part of the same resultant.  The following subsections will discuss each in turn. }

\begin{figure}
    \centering\includegraphics[width=0.5\textwidth]{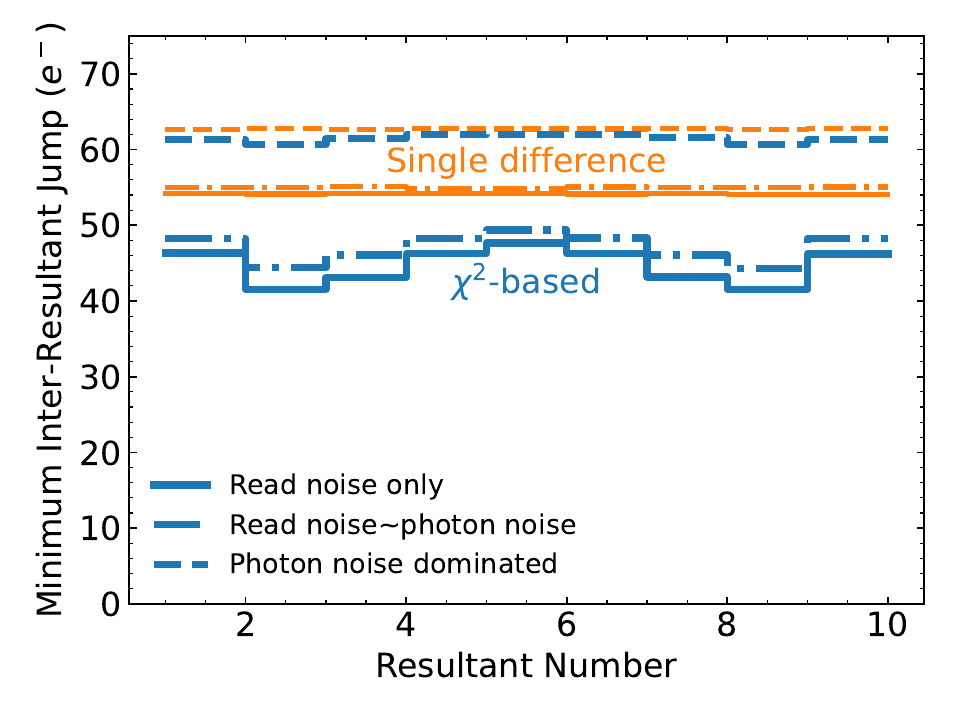}
    \caption{Median sensitivity to a jump that occurs between resultants using either the $\chi^2$ approach presented here (blue steps) or a single resultant difference (orange steps).  The line styles show results assuming only read noise (solid lines), equal photon and read noise when combined over the ramp (dot-dashed lines), and photon noise ten times read noise, again, over the entire ramp (dashed lines).  In the latter case photon noise dominates slightly over read noise in each individual resultant difference. A resultant consists of six reads each with a read noise of 20\,e$^-$. 
    {The $\chi^2$ approach is always more sensitive, though this advantage approaches zero in the limit of pure photon noise.}
\label{fig:sensitivity_resultant_ramp}}
\end{figure}

\subsubsection{{Inter-resultant sensitivity}}

Figure \ref{fig:sensitivity_resultant_ramp} shows the sensitivity of both the $\chi^2$-based approach presented here and a single resultant difference to a jump between resultants.  The figure shows the jump that is detected 50\% of the time assuming a read noise of 20\,e$^-$/read and six reads per resultant.  The equal photon and read noise case corresponds to a count rate of {1.15}\,e$^-$/read; both noise sources then contribute equally to the overall ramp fit.  In the photon noise dominated case (count rate of {11.5}\,e$^-$/read) photon noise dominates over the entire ramp, {but} is slightly {smaller} than read noise in an individual resultant difference.  

{The $\chi^2$ approach is more sensitive than the single difference approach, though this advantage asymptotes to zero in the limit of pure photon noise.  The $\chi^2$ approach must also consider whether to discard one or two resultant differences when it is possible that a jump occurred within a resultant.}  It is only correct to discard one resultant difference if it is known that the jump occurred between resultants; this can be difficult to assess.  
In the $\chi^2$ approach this {choice} is integrated into the process: both single and pairs of excluded differences are tested using user-supplied significance thresholds that can place them on equal footing.  {The following subsection explores the sensitivity to the $\chi^2$ approach to jumps that are injected within resultants.}

\subsubsection{{Intra-resultant sensitivity}}

A jump in the middle of a resultant is more difficult to detect than a jump between resultants.  The additional counts are diluted: the resultant with the jump only contains half of the extra counts that it would have if the jump happened just before the first read of the resultant.  The difference between resultants likewise consists of smaller jumps by up to a factor of 2.  A single-difference search is limited by these facts, while a likelihood approach can do better because it can account for the jump across more than one resultant difference.

Here I compare a single resultant difference threshold with my $\chi^2$-based likelihood approach for a jump that occurs within a resultant.  I again compute the jump level that will be flagged 50\% of the time and choose thresholds that result in consistent false positive rates between the two approaches.  To show the difference in sensitivity as a function of jump time within a resultant, I normalize sensitivities to the inter-resultant sensitivity at which both approaches are most sensitive to jumps.  I exclude the first and last resultants from this analysis.  A jump within the first or last resultant is both more difficult to detect and shows little difference between the $\chi^2$ approach and the single difference approach.  Both of these are because only a single resultant difference is available.

\begin{figure}
\includegraphics[width=0.5\textwidth]{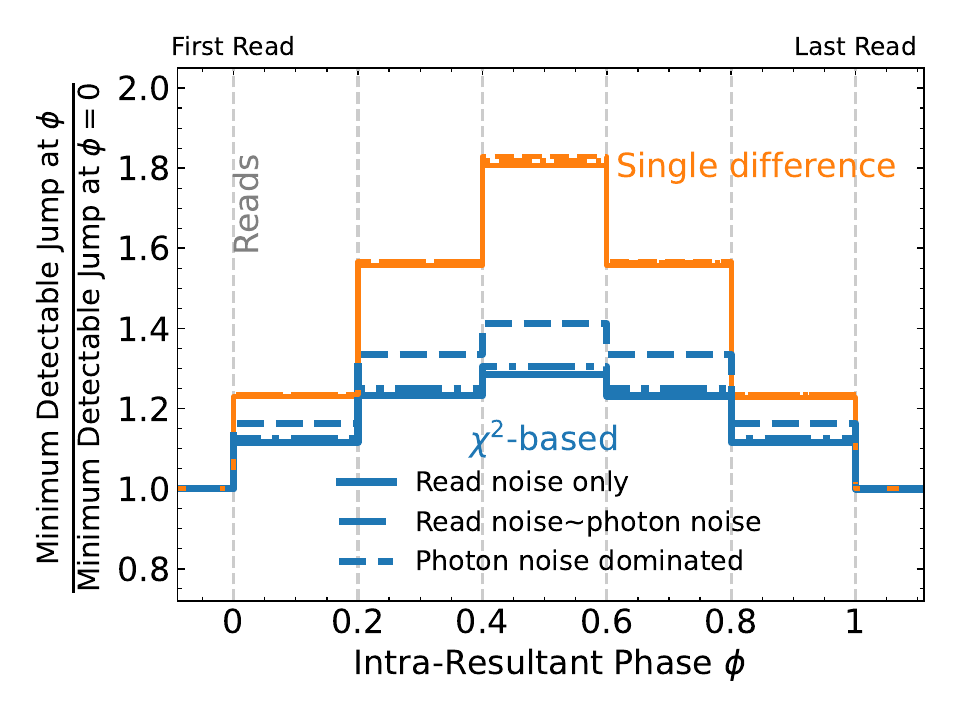}
\includegraphics[width=0.5\textwidth]{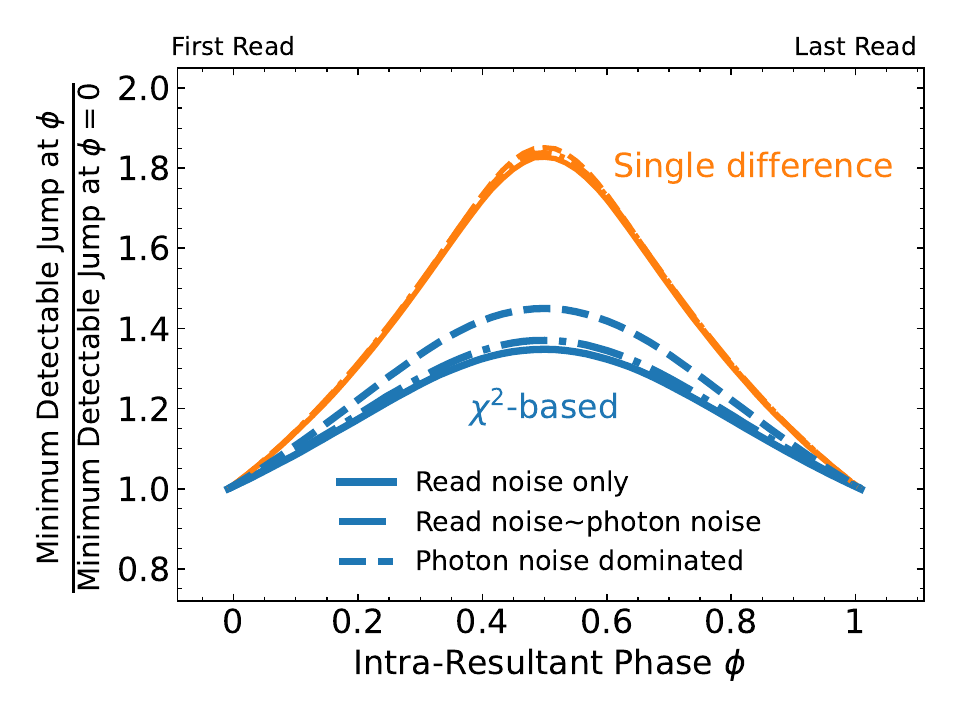}
\caption{Left panel: minimum detectable jump at 50\% efficiency as a function of its position within a 6-read resultant in a 10 resultant ramp; the jump level is normalized to its value between resultants (where efficiency is maximized).  Right panel: the same plot for 10 resultants each composed of a large number of reads (60 in these simulations).  A detection algorithm only using a single resultant difference loses almost half its sensitivity when a jump happens halfway through a resultant.  In this case the peak resultant difference is half of its value for a cosmic ray that hits between resultants.  The median sensitivity drops by slightly less than a factor of two because there is an extra chance to get a favorable realization of noise to add to the jump.  The $\chi^2$ approach described here is substantially more sensitive especially when read noise is important. 
\label{fig:intraresultantanalysis}}
\end{figure}

Figure \ref{fig:intraresultantanalysis} shows the results for six-read resultants on the left and for many-read resultants on the right (60 in these simulations).  The read noise is again assumed to be 20\,e$^-$/read and the count rates are 0, {1.15}, and {11.5}\,e$^-$/read with six reads/resultant, and 0, {0.0115}, and {0.115}\,e$^-$/read with 60-{read} resultants.  The middle count rate in both cases yields an equal contribution of photon and read noise to the overall ramp fit.  

The single-difference approach suffers from almost a factor of two loss in sensitivity for a jump that occurs in the middle of a multi-read resultant. The signal is a factor of two lower than for a jump between resultants, but with two equal differences (one on either side of the resultant with the jump), one is likely to have a noise realization that adds to the difference caused by the jump.  For independent Gaussian noise of variance $\sigma^2/2$ in each resultant (i.e.~variance $\sigma^2$ in each resultant difference), the larger of two adjacent resultant differences has a median of $0.64 \sigma$.  
Assuming a threshold of $4.5\sigma$ this would give a ratio of intra-resultant to inter-resultant sensitivity of 
\begin{equation}
    \frac{2 \times (4.5 - 0.64)}{4.5} \approx 1.71 .
\end{equation}
where the factor of 2 in the numerator is because only half of the jump occurs in each difference.  This effect is partially countered by the fact that the reference count rate must be estimated from the reads themselves.  The median of the scaled read differences is biased high relative to the true count rate because one or two of these differences contain contributions from a jump.  This bias is $\approx$0.3$6\sigma$ when two of nine resultant differences are biased many sigma high by a jump; it is $\approx$0.1$5\sigma$ when only one of nine resultant differences is biased high.  This increases the ratio of the intra-resultant to the inter-resultant sensitivity to slightly more than 1.8, in closer agreement with Figure \ref{fig:intraresultantanalysis}.  Biases in the count rate also produce overestimated noise values and reduce sensitivity, slightly increasing the sensitivity ratio to its value of almost 1.85 in Figure \ref{fig:intraresultantanalysis}.  

The $\chi^2$ approach is significantly more sensitive than a single difference threshold, especially where read noise dominates.  Because the $\chi^2$ approach uses both differences containing the resultant with a jump, it is a factor $\approx\sqrt{2}$ more sensitive than using single-result differences when the resultants have negligible covariance.

The $\chi^2$ approach has an additional advantage over a single difference threshold.  The $\chi^2$ approach omits two resultant differences at a time; it excludes both differences affected by the resultant containing a jump.  In this way it identifies the specific resultant containing the jump: the one that is shared between the two omitted differences.  A single difference threshold identifies a difference containing the resultant with a jump.  However, it does not identify which of the two resultants contributing to this difference contains the jump. A jump near the end of the first resultant looks identical to a jump near the beginning of the second resultant.  As a result, two resultants (three resultant differences) might need to be excluded, rather than one resultant (two differences) with the $\chi^2$ approach. 

\section{Implementation} \label{sec:implementation}

I have implemented the algorithms described in this paper in pure Python; the core of the implementation was described in \citetalias{Brandt_2023}.  The jump detection step is integrated into the same package and uses some of the same functions: the ramp fitting function includes a flag to also search for jumps.  As for \citetalias{Brandt_2023}, all tests running the code were performed on a 2020 Macbook Air, and the code includes tests to verify the results and performance of the jump detection step using jumps added to synthetic data.

The jump detection used is an implementation of the iterative jump detection algorithm described in Section \ref{sec:jumpdetect}.  This function is built on the ramp fitting function described in \citetalias{Brandt_2023}, which can also compute slopes and $\chi^2$ values omitting all resultant differences or pairs of differences in turn.  As noted in Section \ref{sec:jumpdetect}, if a resultant difference is omitted even though it has been previously masked, the equations of Section \ref{subsec:omitone} and \ref{subsec:omittwo} can give 0/0, which evaluates to {\tt NaN}.  All {\tt NaN} values are then replaced with the $\chi^2$ values and slopes corresponding to omitting one fewer resultant difference. 

The outputs from this step include a best-fit count rate for each pixel, its standard error, and a mask array of the resultant differences used.  The latter array is one for each difference used in the ramp fit and zero for each difference not used.  The jump detection algorithm can operate on a mask array that is set to zero for, e.g., saturated pixels.  In this way, the jump detection algorithm strictly reduces the number of valid resultant differences.  Once this mask is computed, the count rates and uncertainties may be computed exactly as described in \citetalias{Brandt_2023}.

The computational cost of jump detection is linear in the number of resultants.  It does depend modestly on the number of jumps present, as this determines how many iterations are necessary.  Running cosmic ray detection takes $\approx$9 seconds per $10^8$ pixel-resultants without, or $\approx$12 seconds per $10^8$ pixel-resultants with, one additional fit to remove bias, assuming that the number of jumps is small compared to the number of pixels (few pixels require multiple iterations).  For an H4RG ramp with 10 resultants this cost corresponds to $\approx$20 seconds to perform a full cosmic ray cleaning.  The time to perform a full cosmic ray cleaning scales modestly with the number of cosmic ray hits. {It increases by $\approx$20\% when cosmic rays impact 1\% of resultant differences (similar to the rate in Section \ref{subsec:ex_deep8} and larger than the rate in Section \ref{subsec:ex_rapid}), and} approximately doubles in the extreme case where cosmic rays affect 10\% of the resultant differences.

\section{Examples Using JWST Data}
\label{sec:examples}

Section \ref{sec:jumpdetect_demo} demonstrated the jump detection algorithm on synthetic data.  In this section I apply it to real data from the NIRCam instrument \citep{Rieke+Kelly+Horner_2005} on JWST (note that JWST uses the term ``group'' instead of ``resultant'').  I fit the raw groups without applying any nonlinearity or bias corrections and use the calibration data products publicly available\footnote{\url{https://jwst-crds.stsci.edu/}}.  I do apply a reference pixel correction, though not the same one as the JWST pipeline.  I subtract the first group from all subsequent groups and then subtract the mean of the reference pixels at the end of each readout channel.  Next, I smooth the pattern given by the reference pixels along the sides of the first and fourth readout channel, and subtract this smoothed pattern from all channels.  A smoothing kernel with a standard deviation just over six columns works well, with 0.94 times the resulting pattern subtracted from each column.  {These numbers are chosen to minimize the $\chi^2$ value of the ramp fits at fixed read noise, i.e., to provide the smallest residuals about the best-fit ramp.  The reference pixel correction I apply works significantly better than the default approach in the JWST pipeline, which uses a median filter rather than a Gaussian kernel to smooth the signal from the side reference pixels.  The Gaussian kernel reduces the effective read noise by $\approx$5\%-10\% below the level with median smoothing. }  I apply no further calibrations before fitting a ramp.

Both data sets I use here are from Detector 1 of NIRCam, the short wavelength detector.  I compare the results of the new algorithms with the publicly available {\tt \_rate.fits} files.  I flatfield both the {\tt \_rate.fits} files and the new fits to remove pixel-to-pixel variations in sensitivity and illumination; this facilitates a straightforward visual comparison.

\subsection{NIRCam RAPID Exposure} \label{subsec:ex_rapid}

I first demonstrate the new algorithms on NIRCam RAPID data with eight reads, and a single read per group\footnote{filename {\tt jw02731001001\_02105\_00004\_nrca1\_uncal.fits}}.  These data are from Early Release Science (ERS) imaging of NGC 3324 in the F200W filter with the {\tt nrca1} detector.  The total exposure time was 161 seconds.  I use the default cosmic ray rejection threshold of $\Delta \chi^2 = 20.25$ corresponding to $4.5\sigma$.

\begin{figure*} 
\includegraphics[width=\textwidth]{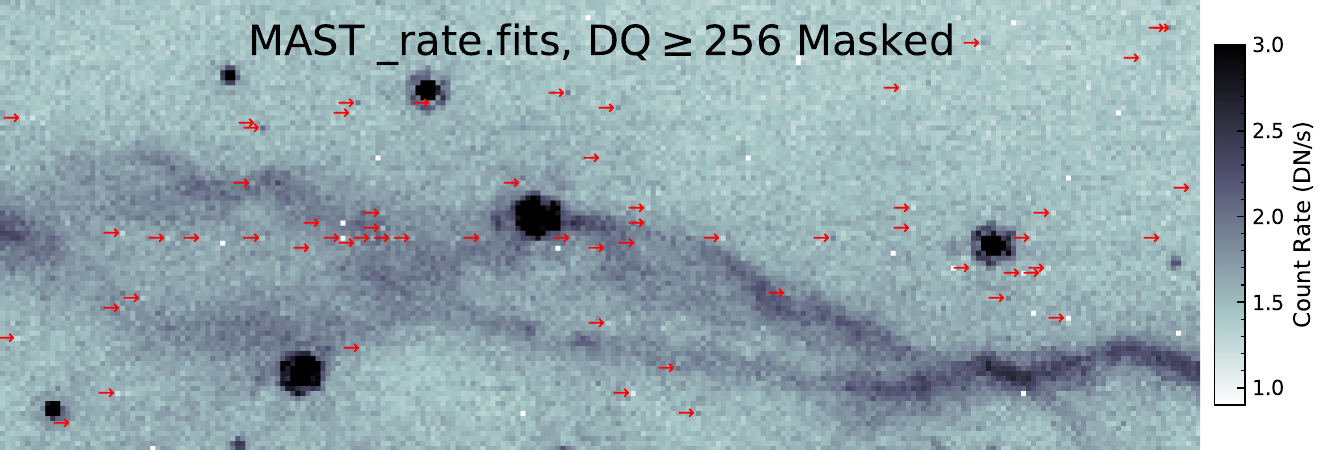}\\[1em]
\includegraphics[width=\textwidth]{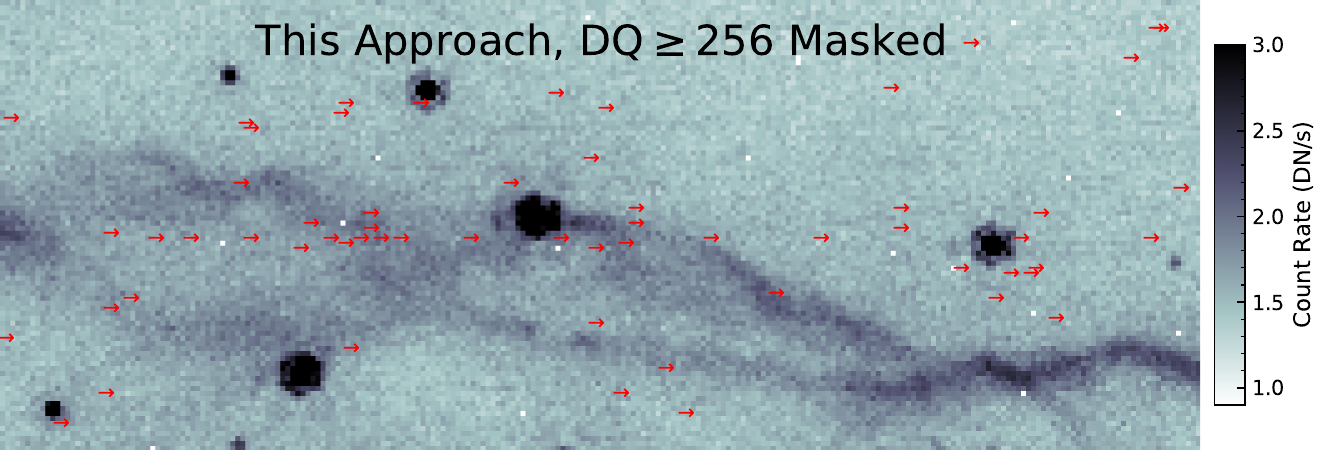}
\caption{Comparison of NIRCam short wavelength images constructed using the the algorithm described here (top) and from the {\tt \_rate.fits} file on MAST (bottom). The data were taken in RAPID mode, with eight reads saved individually.  In both panels the images have been scaled by the flatfield for clarity.  The two images are nearly identical, but close inspection shows a few differences.  {Red arrows indicate good pixels with count rates $<$3\,DN/s for which the two fitted count rates differ by at least 0.2\,DN/s.} Pixels flagged in the calibration files (${\rm DQ} \geq 256$) are masked in both images.
 \label{fig:rapid_comparison}}
\end{figure*}

Figure \ref{fig:rapid_comparison} compares a small region of the {count rate} image in the {\tt \_rate.fits} file available on MAST (left, pipeline version 1.10.1) with my {maximum likelihood count rate image} (right); the color scale is logarithmic.  In both cases I have flatfielded the image.  I have also masked pixels with data quality flags ${\rm DQ} \geq 256$, i.e., the pixels that are flagged based on calibration files rather than the result of the ramp fit.  The same pixels are masked in both the {top} and {bottom} image.  

\begin{figure*}
    
    \begin{center}    
    \includegraphics[width=0.5\textwidth]{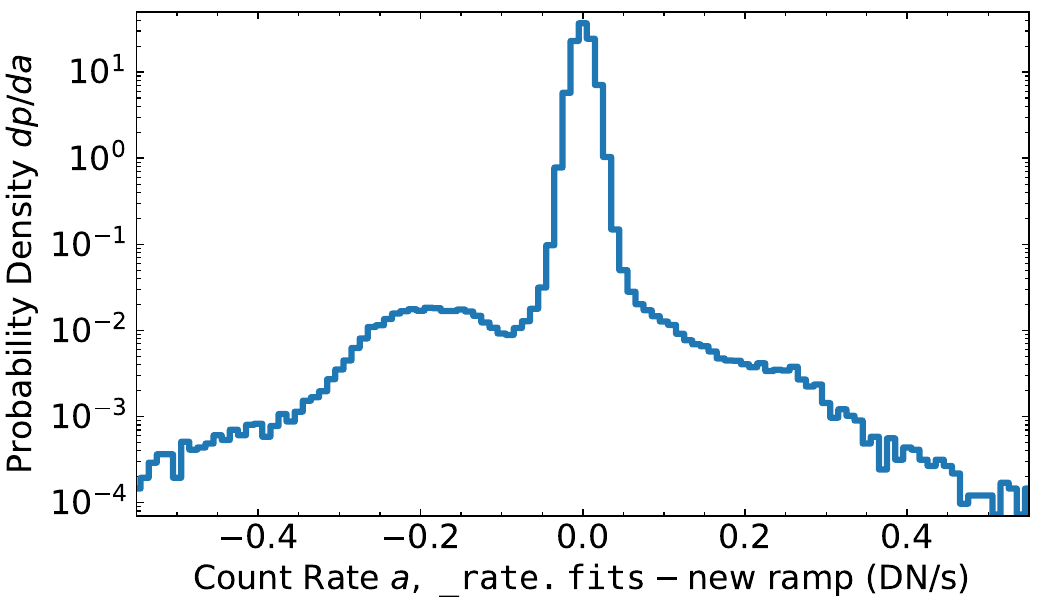}  \\[1em]
    \includegraphics[width=0.5\textwidth]{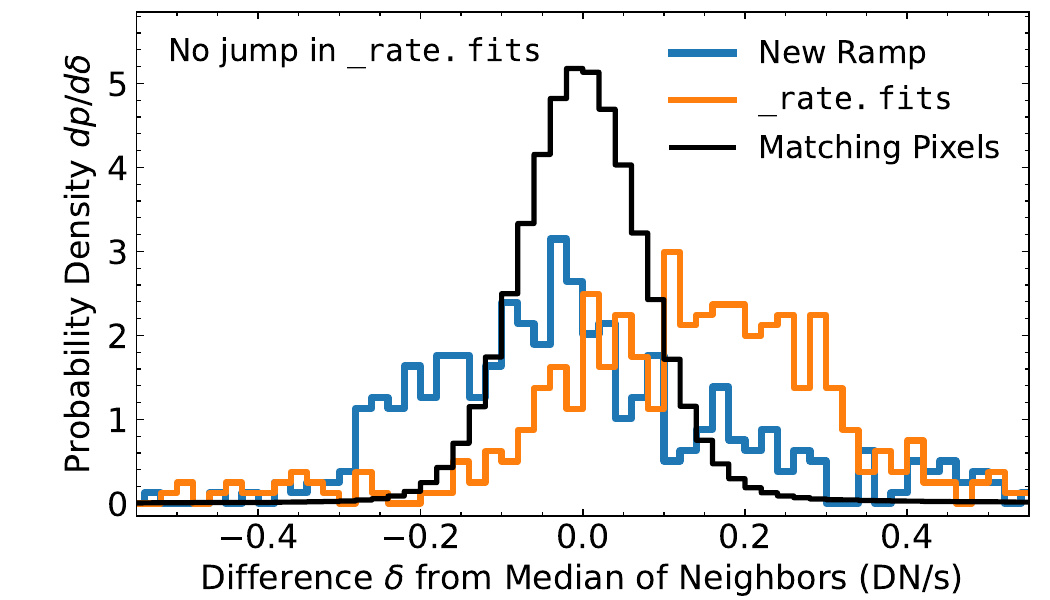}   \\[1em]
    \includegraphics[width=0.5\textwidth]{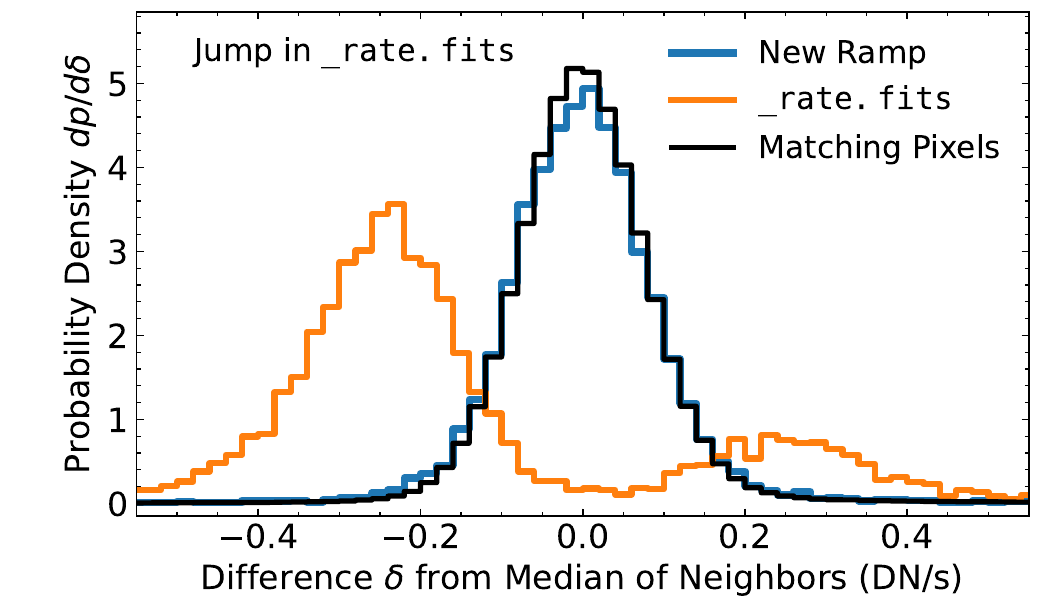}   
    \end{center}
    \caption{{Top:} histogram of the difference in fitted count rate between the MAST {\tt \_rate.fits} {map} and my new {maximum likelihood fit}, restricted to good pixels far from nonlinearity (${\rm DQ} < 256$, DQ bit 1 is zero, $a < 5$\,DN/s); note the log scale.  The great majority of pixels have a negligible difference, but $\approx$0.3\% of pixels differ by more than 0.2\,DN/s. {Middle and bottom:} differences between a pixel's value and the median of its neighbors for pixels that agree to 0.2\,DN/s between the two {fitted count rates} (black histograms), and for the pixels that differ by at least 0.2\,DN/s between the two {count rate maps} (blue and orange histograms). {The middle panel shows pixels where a jump is found in the new reduction but not in the {\tt \_rate.fits} file; the bottom panel shows pixels where a jump was flagged in the {\tt \_rate.fits} file but not in the new reduction.} \label{fig:discrepantpixels}}
\end{figure*}

The two images in Figure \ref{fig:rapid_comparison} look nearly identical, but a close look shows a few pixels that are discrepant.  {Small red arrows highlight pixels with a count rate $<$3\,DN/s for which the two fitted count rates differ by at least 0.2\,DN/s. }  
Figure \ref{fig:discrepantpixels} provides a closer look at {the discrepant} pixels.  The {top} panel shows a histogram of the difference in the flatfielded pixel values between the MAST {count rate} and my {maximum likelihood count rate}.  The vast majority of pixels have a negligible difference, but a fraction of a percent show strong discrepancies.  
The {middle and lower} panels of Figure \ref{fig:discrepantpixels} show the difference between a pixel and the median of its neighbors, both for the pixels that match to within 0.2\,DN/s between the two {count rate maps} (black histogram) and those that do not (orange and blue histograms).  The black histogram shows the agreement between a pixel's {count rate} and {that of} its neighbors that we should hope to see for a good fit {to the ramp}.  The distribution will be broader for noisier pixels or pixels that must be fit using only some of the resultants.  {About 0.3\% of pixels are discrepant between the two ramp fits.  The middle panel of Figure \ref{fig:discrepantpixels} shows pixels for which the {\tt \_rate.fits} file did not detect a jump but my new reduction did, while the lower panel shows pixels where the {\tt \_rate.fits} file identifies a jump where my reduction identifies none. The bottom panel contains slightly more ($\approx$20\%) pixels than the middle panel.  }

The pixels that differ between the two {count rate maps} show very different properties in the different {ramp fits}.  {For the pixels where the maximum likelihood fit identifies a jump that the {\tt\_rate.fits} map does not, the pixels in both maps show large deviations from their neighbors, with the {\tt\_rate.fits} pixels showing a positive bias.  For the pixels where the maximum likelhood fit does not identify a jump that the {\tt \_rate.fits} file does, the pixels show much better agreement with their neighbors in the maximum likelihood count rate map.  The discrepant pixels in the MAST count rate map (orange histogram), in contrast to those in the maximum likelihood map, show poor agreement with neighbors. This is strong evidence that these jumps were identified incorrectly in the MAST images. }

\subsection{NIRCam DEEP8 Exposures}
\label{subsec:ex_deep8}

I next demonstrate the new algorithms on a sequence of seven integrations taken with Detector 1 of the NIRCam instrument on JWST, proposal ID 2079\footnote{filename {\tt jw02079004001\_03201\_00001\_nrca1\_uncal.fits}}.  Each integration has four groups taken in DEEP8 mode, with eight reads averaged per group, and twelve reads skipped between groups.  The first read is saved separately from the first group (it is also present in the first group), but this separate read is not used by the JWST pipeline.  This data set poses different challenges from the previous one.  There are fewer groups, the groups have unequal numbers of reads if we are to use the first read separately, and cosmic rays will impact the ramps differently if they arrive within rather than between groups.  

The JWST pipeline currently treats each {integration} separately.  This is unable to effectively flag jumps with just four groups.  If a jump occurs within the second group, for example, the differences between Groups 1 and 2 and between Groups 2 and 3 are both corrupted.  Only the difference between Groups 3 and 4 is a reliable indicator of the {count rate}.  But because there are only three group differences available, there is no way of knowing that the difference between Groups 3 and 4 is the one to use, unless we place a prior that jumps should be positive and take the smallest group difference as the valid one.  {This provides a significant incentive to make full use of the first read.}   

To process the data, I first separate out the first read of each integration.  Since the first group {\tt GROUP1} is the average of the first eight reads, I set the new first group (which I will treat as the second group, {\tt GROUP2}) to
\begin{equation}
    {\tt GROUP2} = \frac{8}{7} \left({\tt GROUP1} - {\tt READ1}/8\right).
\end{equation}
The new {\tt GROUP3} will be identical to the old {\tt GROUP2}, etc., while the new {\tt GROUP1} is the old and previously ignored {\tt READ1}.  

Next, I arrange all groups from all integrations sequentially, resulting in 35 groups, or 34 differences between groups.  There is a reset between {integrations}, e.g., {between} Groups 5 and 6, which I treat as a jump; these group differences will be ignored by the ramp fit.  This leaves 28 group differences to be used in fitting a ramp.  {By combining all integrations at the group level, and by separating out and fully utilizing the first read, we can slightly improve our signal-to-noise ratio and regain sensitivity to jumps that the JWST pipeline cannot detect.  }

I compute the $\alpha$ and $\beta$ components of the covariance matrices as described in \citetalias{Brandt_2023} and perform a fit including jump detection.  To better account for cosmic ray ``snowballs'' and persistence after strong hits, if three consecutive resultant differences have a jump detected, I discard a further three after the last detected jump.  ``Snowballs'' manifest as a jump over a nearly circular region of the detector \citep{Cillis+Cottingham+Waczynski+etal_2018}.  If there is a large ($\gtrsim7\times 7$) contiguous region with a detected jump in the same group difference, I measure the radius of the contiguous region and extend the jump mask to a region 2.5 times this radius in the affected group difference.  An expansion factor of 2.5 was the smallest that removed all traces of the snowball to my eye.

\begin{figure*} 
\includegraphics[width=0.5\textwidth]{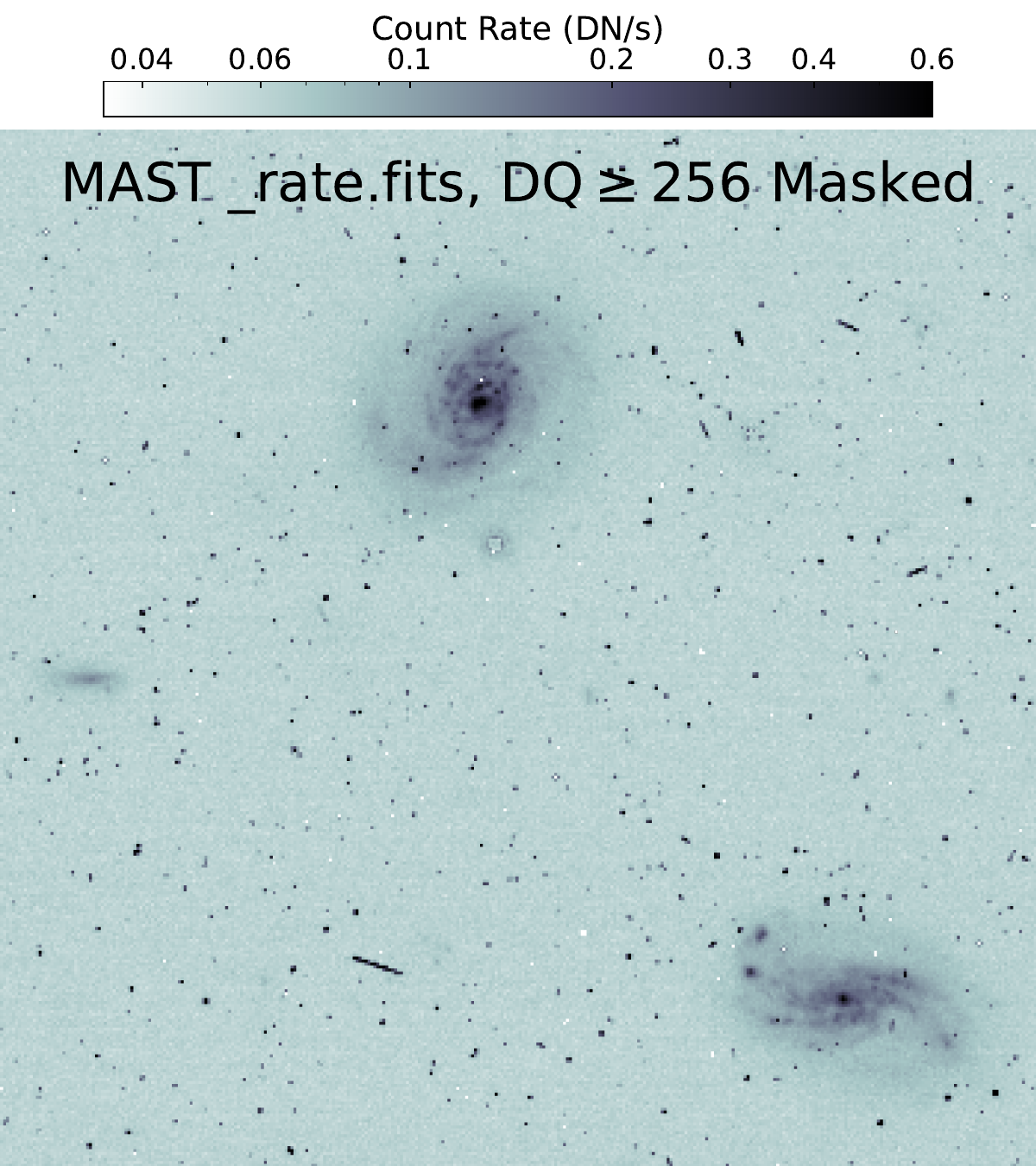}
\includegraphics[width=0.5\textwidth]{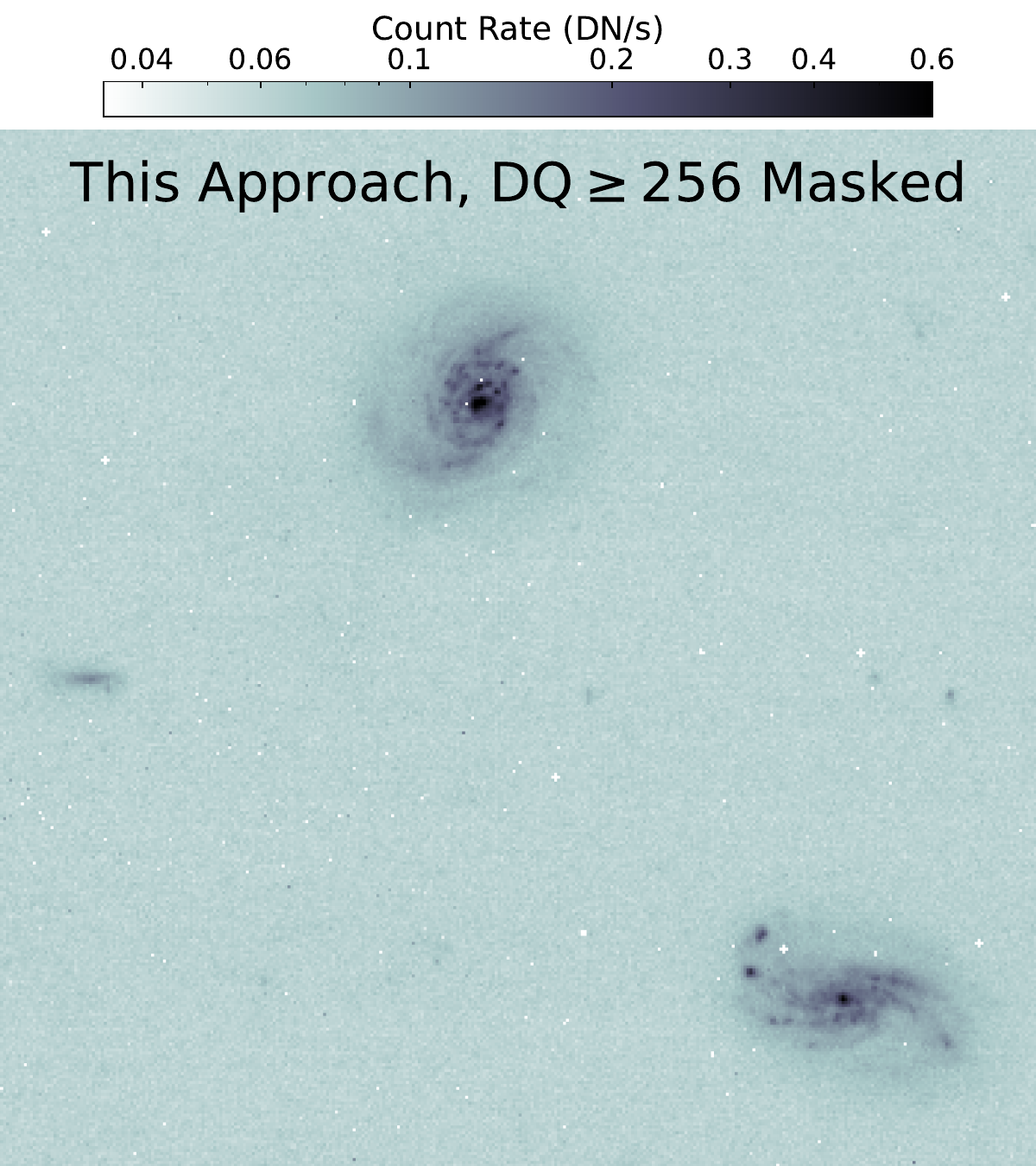}
\caption{Comparison of a subregion of the {\tt \_rate.fits} file available on MAST as of November 2023 (left, pipeline version 1.11.4) with the approach presented in this paper (right) for a DEEP8 integration with NIRCam in the short wavelength channel.  In both cases, pixels are masked if they are flagged in the calibration files rather than only in the fit to the ramp (${\rm DQ} \geq 256$).  In the MAST file, 48\% of non-reference pixels have at least one jump detected; the corresponding figure is 30\% in the approach presented in this paper. 
 Both images have been flatfielded for visual clarity and use a logarithmic color scale. \label{fig:deep8_comparison}}
\end{figure*}

Figure \ref{fig:deep8_comparison} shows a small region of the {\tt \_rate.fits} file available on MAST (pipeline version 1.11.4) alongside the {count rate} that I measure.  In both cases, I have scaled the {count rate maps} by the relevant flatfield for visual clarity.  I have also masked pixels flagged by the calibration files (e.g.~hot pixels, warm pixels, dead pixels, telegraph pixels, etc.) but not pixels flagged as jumps or saturations by the ramp fit itself.  These are indicated by a data quality, or DQ value, of at least 256. 
The MAST image has nearly 48\% of pixels flagged as having a jump detected, yet the {count rate} image shows clear indications that these jumps are not being effectively removed.  The {maximum likelihood count rate image} has just $\approx$30\% of pixels with a jump detected in at least one group difference, but the final {count rate} image shows that nearly all cosmic rays and artifacts have been removed.  I can compute the penalty in signal-to-noise ratio due to these jumps by comparing the computed noise with these corrupted group differences masked to the noise that would be derived if all group differences were valid.  The penalty in signal-to-noise due to the loss of information from jumps is typically $\lesssim$5\%.

\begin{figure}
    \includegraphics[width=0.5\textwidth]{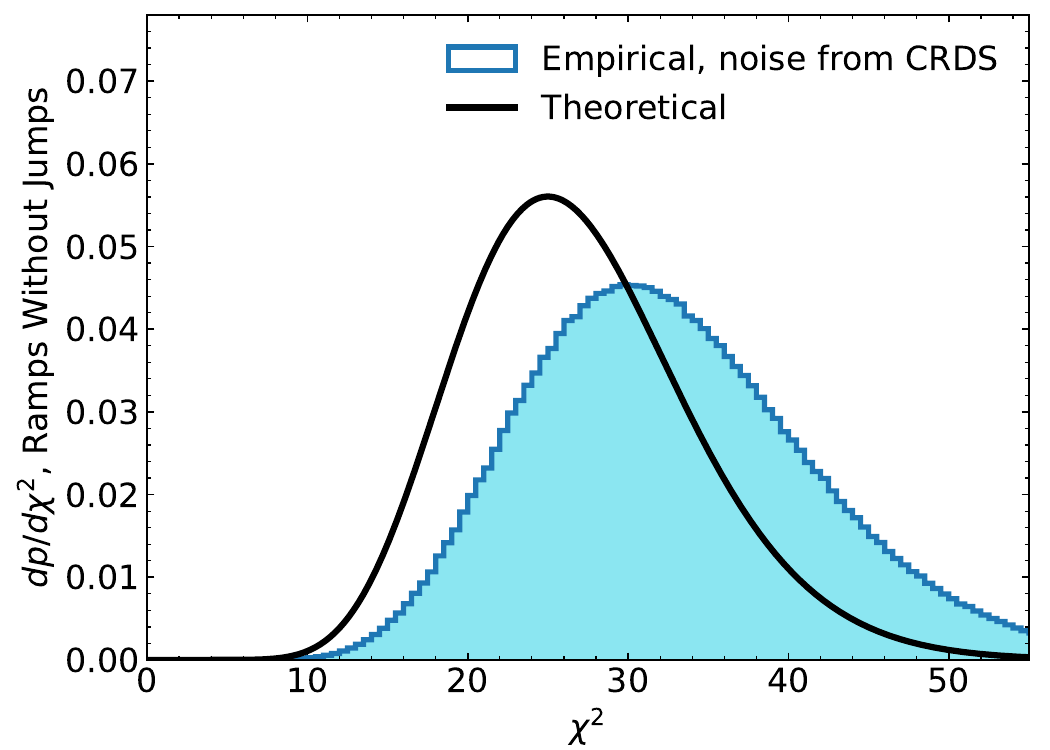}
    \includegraphics[width=0.5\textwidth]{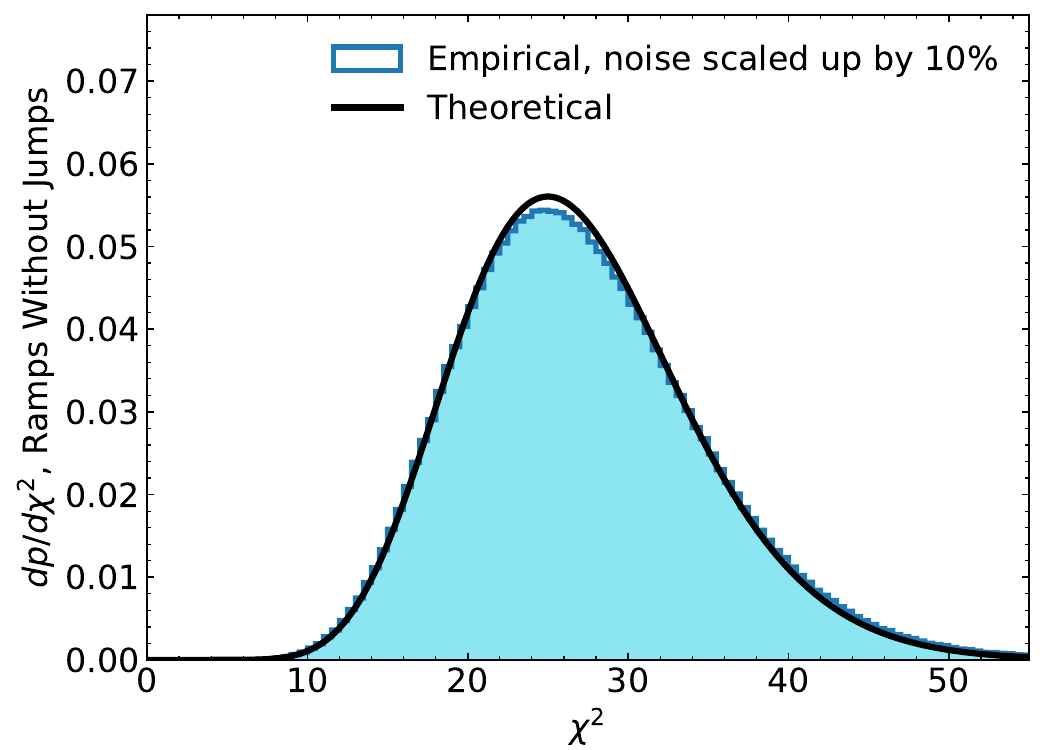}
    \caption{Histogram of the $\chi^2$ values of the DEEP8 ramp fits shown in Figure \ref{fig:deep8_comparison}, restricted to the $\approx$70\% of pixels with no detected jumps. The empirical histogram matches the theoretical $\chi^2$ distribution well if the noise (both read and photon noise) is inflated by 10\% (right panel), but the $\chi^2$ values are systematically high adopting the noise from CRDS.  The theoretical curve has 27 degrees of freedom: 28 group differences minus one fitted slope.  These results suggest that the true uncertainty in the slope should be slightly larger than the formal uncertainty derived from the calibration file's gain and read noise.}
    \label{fig:chisq_deep8}
\end{figure}

Figure \ref{fig:chisq_deep8} shows the $\chi^2$ values for the ramp fit for the $\approx$70\% of pixels without detected jumps.  The comparison curve is a $\chi^2$ distribution with 27 degrees of freedom: 28 group differences minus one fitted slope.  The agreement is good if I inflate the uncertainty implied by the calibration file's gain and read noise by 10\% (i.e.~if I divide $\chi^2$ by 1.21).  This suggests that the adopted covariance matrix remains a good overall description of the data, but that there are sources of noise in long exposures with many reads that are not fully captured by the calibration files.  Figure \ref{fig:chisq_deep8} suggests that the formal uncertainties on the count rates may need to be inflated by $\approx$10\%.  If some of this noise can be removed by, e.g., a better suppression of $1/f$ noise \citep[e.g.][]{Moseley+Arendt+Fixsen+etal_2010,Rauscher_2015}, then the necessary inflation factor may be lower.  {I note that I did not use the default JWST reference pixel correction for Figure \ref{fig:chisq_deep8}; the JWST pipeline by default uses a median filter of length 11 to smooth the rows of reference pixels on the sides before subtraction.  Instead, I use the median to combine the eight reference pixels for each row, and then convolve the row-by-row results with a Gaussian as a low-pass filter.  The median filter performs significantly worse on the DEEP8 images than the Gaussian filter that I use.  With a median filter of length 11 instead of a Gaussian filter, I would require inflating the noise by $\approx$15\% rather than $\approx$10\% to obtain good agreement with the theoretical $\chi^2$ distribution. }

Processing the DEEP8 ramp involves reading in the data, performing a reference pixel correction, fitting a ramp including jump detection, applying some additional masking to remove most of the effects of snowballs and persistence, and then fitting the ramp twice again to remove bias.  This entire process takes about 45 seconds on my 2020 Macbook Air, making it practical to apply to large data sets.

\section{Conclusions} \label{sec:conclusions}

This paper uses the ramp fitting approach described in \citetalias{Brandt_2023} to implement likelihood-based jump detection using the full information contained in a ramp.  The approach is computationally efficient; its cost remains linear in the number of resultants in a ramp.  A likelihood-based approach can be substantially more sensitive than a jump search that only compares pairs of resultants, especially for long ramps and for jumps that occur between reads within a resultant.  

I show that the new ramp fitting and jump detection algorithms can offer major improvements in data quality to JWST data taken in readout modes not well supported by the current pipeline.  Even for data that are currently processed well, the new approaches can offer measurable gains.  More testing and validation will be necessary to verify and quantify these gains across JWST instruments and observing modes.

The ramp fitting and jump detection algorithms described here and in \citetalias{Brandt_2023} are implemented in pure Python.  They are publicly available together with a demonstration Jupyter notebook showing how to implement them.  Jump detection is a factor of a few more expensive than ramp fitting, but it remains straightforward on a modern laptop computer, generally taking less than a minute.  The algorithms presented here could be applied to future instruments like WFI on the Roman Space Telescope, or to reprocess archival data from any instrument with a detector read out up-the-ramp.

\software{scipy \citep{2020SciPy-NMeth},
          numpy \citep{numpy1, numpy2},
          Jupyter (\url{https://jupyter.org/}).
          }

\begin{acknowledgements}
I thank Stefano Casertano and Eddie Schlafly for helpful input and suggestions, and Sanjib Sharma, Michael Regan, and Karl Gordon for useful conversations. 
\end{acknowledgements}

\bibliography{refs}{}
\bibliographystyle{aasjournal}

\end{document}